# Impact of lockdowns and winter temperatures on natural gas consumption in Europe

Philippe Ciais[1], François-Marie Bréon[1], Stijn Dellaert[2], Yilong Wang[3], Katsumasa Tanaka[1,4], Léna Gurriaran[1], Yann Françoise[5], Steven Davis[6], Chaopeng Hong[6], Josep Penuelas[7], Ivan Janssens[8], Michael Obersteiner[9], Zhu Deng[10], and Zhu Liu[10]

1. Laboratoire des Sciences du Climat et de l'Environnement, IPSL CEA CNRS UVSQ 91191 Gif sur Yvette Cedex France.
2. Department of Climate, Air and Sustainability, TNO, P.O. Box 80015, 3508 TA Utrecht, the Netherlands.
3. Key Laboratory of Land Surface Pattern and Simulation, Institute of Geographical Sciences and Natural Resources Research, Chinese Academy of Sciences, Beijing, China.
4. National Institute for Environmental Studies (NIES), Tsukuba, 305-8506 Japan
5. Division climat-énergies économie circulaire, Agence d'écologie urbaine, Ville de Paris, Paris, France.
6. Department of Earth System Science, University of California, Irvine, 3200 Croul Hall, California, USA
7. CREAF, Cerdanyola del Vallès, 08193 Catalonia, Spain.
8. Research Group Plants and Ecosystems (PLECO), Department of Biology, University of Antwerp, B-2610 Wilrijk, Belgium.
9. International Institute for Applied Systems Analysis (IIASA), Ecosystems Services and Management, Schlossplatz 1, A-2361 Laxenburg, Austria.
10. Department of Earth System Science, Tsinghua University, Beijing 100084, China.

**Abstract**

As the COVID-19 virus spread over the world, governments restricted mobility to slow transmission. Public health measures had different intensities across European countries but all had significant impact on people's daily lives and economic activities, causing a drop of $CO_2$ emissions of about 10% for the whole year 2020. Here, we analyze changes in natural gas use in the industry and built environment sectors during the first half of year 2020 with daily gas flows data from pipeline and storage facilities in Europe. We find that reductions of industrial gas use reflect decreases in industrial production across most countries. Surprisingly, natural gas use in buildings also decreased despite most people being confined at home and cold spells in March 2020. Those reductions that we attribute to the impacts of COVID-19 remain of comparable magnitude to previous variations induced by cold or warm climate anomalies in the cold season. We conclude that climate variations played a larger role than COVID-19 induced stay-home orders in natural gas consumption across Europe.

**Introduction**

Natural gas use accounts for ≈ 38% of the fossil fuel $CO_2$ emissions in Europe, including the EU27 and UK. The share of this fossil fuel in the energy mix has been growing since 1990 at the expense of oil and coal, as it is viewed as a cleaner source of energy. The main sectors using natural gas are the built environment (residential, public and commercial buildings), the industry, and the electricity production. In addition to steadily growing natural gas use, the COVID-19 pandemic profoundly disrupted economic activities and the energy demand, beginning in early March of 2020. Specifically, the government policies restricted mobility and confined people at home during "lockdowns," and reduced or closed industrial and non-essential commercial activities, with impacts on gas use and pertaining $CO_2$ emissions. Although such impacts have been noted by recent reports [1,2], they have not yet been quantified and analysed in detail.

During the cold season period when the pandemic arrived in Europe, natural gas consumption is typically high due to the heating demand in buildings, which represents 52% of the total gas consumption from all-sectors averaged over January to June. For both residential and commercial buildings, gas consumption is typically inversely proportional to daily air temperature when temperatures fall below a comfort threshold "critical temperature".

Therefore, the lockdowns occurred during a decreasing trend of the heating demand, as temperatures increased from winter to spring. One interesting research question is to quantify and understand the sign and magnitude of natural gas consumption changes, given the opposing effects of people's confinement in households (likely more gas use), the closure of commercial and public buildings (likely less gas use) and the industry temporary shut-down (likely less gas use). Given different national characteristics and different reductions of industrial activities between European countries during lockdowns, a second interesting question is how industrial gas use was affected in 2020. Natural gas use variations in the power sector have been addressed by refs. ($^{3,4}$) and will not be discussed here.

Changes in natural gas consumption during the COVID-19 pandemic translate into roughly proportional changes of $CO_2$ emissions from the combustion of that fuel. Overall, across all fuels and sectors, the reduction in socio-economic activities due the pandemic caused a strong decrease of $CO_2$ emissions in Europe, and the globe. Two recent estimates of the reduction of $CO_2$ emissions during 2020 have been produced, using activity data, by refs. (5,6). In both studies, $CO_2$ emissions from the built environment were the most poorly constrained of all sectors because no direct activity data was available to estimate them. Lacking natural gas consumption data for buildings, Le Quéré et al. made the simple assumption that buildings emissions were reduced in every country according to observed residential use of electricity in the United Kingdom under different levels of confinements. Furthermore, the focus of their study was to estimate the fraction of emissions reductions attributable to COVID-19, not the total emission changes due to COVID-19 and other factors. The contribution of climate factors was thus ignored, even though climate and weather variations strongly influence the heating demand and emissions from gas. In contrast, Liu et al. (2020) estimated daily residential $CO_2$ emissions changes during 2020 by assuming that only climate variables were influential. They used daily air temperature data from ERA5$^7$ with pre-established temperature functions from ref. ($^8$) to calculate daily residential emissions in 2020, disaggregating annual baseline emissions from the EDGAR database $^9$ to daily residential emissions with population-weighted heating degree days in each country.

Here, we analysed daily natural gas use changes from January to June 2020 in Belgium, France, Germany, Italy, the Netherlands, Poland, Spain and the UK (Methods). We collected and harmonized data separated into the power generation, the industry and buildings, also called 'distribution'. In addition, we collected specific natural gas use data for the city of Paris

(Methods). We attempt to analyse whether there has been an increase of gas consumption in the built environment due to a larger number of people being confined or a decrease due to closure of commercial buildings, and to separate the effect of temperature variations from those of socioeconomic activity changes.

**Natural gas consumption variations influenced by temperature**

As a first step, we analysed daily gas use within the city of Paris, which reflects only residential, public and commercial buildings. Paris is an easy case with the same temperature exposure and rather homogeneous housing types, dominated by collective buildings equipped with centralized gas heaters. Figure 1 illustrates the relative impact of temperature anomalies and lockdown measures. Figure 1a (left) shows the consumption as a function of the reference climatological temperature of each day during the period from March 1st to June 1st, when confinement occurred in 2020. The reference temperature increases during the period chosen, and can be seen here as a proxy of the date. March 1st is on the left while the May data points are on the right. At the very beginning of the period (left-most data points), the 2020 consumption appears larger than that of 2019. Yet, later on, in particular during the confinement period, the 2020 consumption appears lower than that of 2019, which could be interpreted as an impact of COVID-19 reducing the gas consumption. However, Figure 1b (right) shows a different perspective: when the same data is shown as a function of the effective temperature, i.e. normalized to remove the effect of daily temperature fluctuations (see below) the difference between 2019 and 2020 consumptions is reduced and the scatter of the data-points as well. These two figures illustrate clearly the important role of the temperature to drive short term variations of gas use, and the need to account for temperature anomalies when looking for the impact of another predictor like the COVID-19 lockdowns severities. For the city of Paris, we conclude that there is no clear impact of the COVID-19 lockdown on natural gas consumption, despite most of the population being confined at home under a very strict confinement.

We then analysed daily gas supply data at national scale (Methods) for eight countries that provided separate estimates for different sectors. Compared to the Paris case, one additional difficulty is the definition of a temperature that is representative for each country. For this, we used the ERA-5 re-analysis [7] and calculated a population-weighted average of air temperature over each country's area, accounting for a temporal-lag impact of temperature:

the consumption is not only dependent on the temperature of the day, but also to that of the previous days (a building has some heat capacity). Namely, we computed an effective temperature based on the temperature at days D-2, D-1 and D-0, with respective weights 0.55, 0.30 and 0.15, which offers the best correlation between effective temperature and gas consumption.

In all the countries, daily gas use shows a decreasing trend with increasing temperature in each country, up to a temperature $T_{crit}$ above which there is no apparent sensitivity to temperature. Examples of this behaviour are shown on Figure 2 for the industry and the buildings sectors in France. For the buildings sector, it is interesting to see in Figure 2 that the consumption during week days is larger than during weekends (Table 1). There is little difference on slope of the T-consumption relationship for France, but the component independent of the temperature is very significantly different between week-days and weekends. This is counter-intuitive as one could expect a larger residential consumption during the weekend when people spend more time at home. A likely explanation is that the distribution data include commercial and public administration buildings that are closed during Saturdays and Sundays. The weekday-weekend difference is rather small for Germany, but goes up to ≈50% for France (Table 1). The industry sector shows a number of "low" outliers for France in Figure 2, most of which are for 2020 and they all occur during the lockdown period, according to Oxford Covid-19 Government Response Tracker data[10] (OxCGRT). Similar negative outliers are observed for the industry in Italy and Spain, but Poland, Germany and UK show positive outliers in the first half of 2020 (Fig S1). On the basis of these figures, we devised piecewise linear models to describe the consumption-temperature relationships. The models were fitted against 2016-2019 daily data points with three parameters: $T_{crit}$, the base gas consumption for temperature above $T_{crit}$, and the slope of the linear relationship between temperature and gas consumption for temperatures below $T_{crit}$.

**Countries differences in the temperature sensitivity of natural gas consumption**

There is a high correlation ($R^2 > 0.9$) between the actual consumption and the one which is predicted by our piecewise linear models, indicating that air temperature is a very good predictor of short-term gas consumption anomalies during the cold season, despite the large range of mean winter temperature between southern and northern countries. Across the eight countries analysed, the critical temperature $T_{crit}$ obtained from the piecewise linear models is remarkably stable between weekdays, Saturdays and Sundays (a few tenths of a degree differences) and varies from ≈14° in the UK to ≈16° in Italy with a broad North-South gradient (Table 1). Conversely, the correlations for the industry sector are much smaller, indicating that temperature is not the only predictor of industrial gas consumption variability, and that there are country differences (Fig S1). In Poland, the Netherlands and the UK, the slope of the industrial gas consumption v.? temperature tends to be smaller than in other countries (Fig S1).

We found that the per capita mean slope (2016-2019) of gas consumption vs. temperature which defines the temperature sensitivity of gas use in the distribution sector, differed between countries, reflecting different building energy efficiency, habits of people, and possibly prices of natural gas. The base (independent of temperature) per capita consumption varies between 0.37 and 0.93 $m^3$ $cap^{-1}$ $day^{-1}$, although Poland is an exception with much smaller values (Table 1). The temperature-dependant sensitivity varies between 0.16 and 0.39 $m^3$ $cap^{-1}$ $day^{-1}$ $deg^{-1}$, again with Poland outside of this range. Poland appears to be an outlier because it has a smaller share of dwellings and population using gas as a heating source ; Gas represents only 7% of the heating consumption in terms of final energy use in Poland, against 41% for coal, 20% for biomass and 29% of district heat [11]. The temperature sensitivity of natural gas use per capita should be directly related to the energy need for residential heating. Using building average national data from different databases on energy use (Methods) and gas heated floor area, we found a strong positive linear relation with the gas used temperature sensitivity and the per capita average gas heated floor area (Figure 3). The country with the largest per capita gas heated floor area is the Netherlands, and the smallest is Poland. The strong correlation shown in Figure 3 implies that the temperature sensitivity of natural gas use per capita depends primarily on the heated floor area per capita, in other words that country differences in insulation for gas heated buildings are of secondary importance. Using detailed subnational data at city or district level instead of national data as in this study, may reveal that other factors control the temperature sensitivity of gas consumption, like the types of building (collective vs. individual) and their insulation characteristics / energy efficiency which generally decrease with building age. It is also possible that other meteorological variables such as humidity, surface radiation, and sunshine duration play an additional role in determining the amount of natural gas use in the distribution sector.

**Changes between the lock-down and pre-lockdown periods**

We first look at changes of consumption in the distribution sector. The winter 2019-2020 has been exceptionally mild over Europe, with a warming anomaly of ≈ 1°C from December to March, compared to the climatology of our study period [7]. At the beginning of lockdowns however, between mid-to-late March and early-April, a succession of colder than normal weather events occurred, bringing cold spells over western Europe. When removing the effect of these cold spells using the piecewise linear models described above, we display in Figure 4 the relative changes of buildings' gas consumption in France that are not explained by temperature during the lockdown period (March $1^{st}$ to June $1^{st}$) compared to the pre-lock-down period (January $15^{th}$ to March $1^{st}$). In Italy, France, Belgium, Germany and Poland, we found a relative reduction, larger than one standard deviation of variations between the two periods during other years. On the other hand, in the UK and the Netherlands, there is a small relative increase during the lockdown periods as compared to the pre-lockdown reference period, but this change remains within the one standard deviation of other years. This difference might be explained by the large share of houses heated by natural gas in UK and Netherlands, and by less stringent lock-down measures that allowed a fraction of people to continue to remain in workplaces. This result indicates that despite nearly all the population being confined at home all day long in Spain, Italy and France, and the strong mobility reduction, e.g. by 75% in France[12], the natural gas consumption rather decreased than increased. This may be due to some houses being empty compensating for other full households, and to closed commercial and public buildings with reduced or no heating.

In the seven countries where building consumption data was available, we regressed daily anomalies of gas consumption corrected for temperature against the OxCGRT government response indexes, and the Google mobility indexes (relative presence at home) for the period Mar 1-Jun 1, 2020. Significant but weak negative correlations were found for all countries, except in Germany and UK, which had less strict confinements than other countries (Fig. S2). These correlations between government response indexes (lockdowns severity) and gas consumption are impaired by the sparsity of data with lockdowns lasting less than 30 days during the cold season when gas is used for heating, and by the fact that OxCGRT indexes take discrete values and stall to zero when lockdowns are released. Regressions with Google mobility data were also weak, but and they show the same negative correlation sign than with OxCGRT data. These generally negative correlations suggest that overall, empty commercial and administrative buildings accounted a drop of gas consumption larger than any increase from fully occupied households. Our data did not separate commercial from residential buildings to prove this hypothesis, but an analysis of gas use data in the US confirmed that most of the reductions in gas consumption came from the commercial sector[13].

The industry sector gas consumption is not as temperature-dependent as the distribution sector is. Nevertheless, a temperature correction was applied from linear piecewise models fitted to daily consumption data from the industry sector (Methods) to properly identify the consumption anomalies that are linked to the lockdown. The time series of industry gas consumption is shown in Figure 4 for France as an example. On this figure, the impact of the lockdown is very clear, with a strong and rapid decrease of the consumption just after the start of the lockdown. The consumption returned to near normal values after about two weeks, once adequate procedures had been put in place in the industries.

Figure 5 summarizes the inter-annual variations of the gas consumption during 2020 and previous years, for both the industry and the distribution sectors. Consumption relative deviations from normal values (2016-2019 mean) are shown. The figure also shows the variation between "pre-lockdown" and country specific "lockdown" periods defined from ref.[14]. For the industry sector, the largest impact is over Italy, with a 25% decrease in consumption. The lockdown impact is also very clear for this sector in France, Spain, Italy and Poland. For UK and the Netherlands, there was no decrease of gas use by the industry during the lock down, consistent with the fact that in May 2020, these two countries had the smallest decrease of industrial production among the 8 countries considered [15]. As for the building sector, most countries show a small decreased consumption of 10 to 20%, but there is no such signal over Germany and the UK, where mobility was less restricted than in the other countries studied.

**Conclusions**

Our results show that the largest impact of COVID-19 response measures natural gas use was on the reduction of natural gas consumption by the industry sector in Italy (25%), France (16%), and Spain (14%), but also considerable and shorter duration reductions in Germany and Poland. Meanwhile, the lockdowns also caused a net decrease of gas distribution to the built environment, by roughly 15% in each of Italy, Belgium, the Netherlands, and Poland. However, the anomalies in building's gas use due to COVID-19 lockdowns (after removing the effect of temperature), are of the same magnitude as the anomalies related to temperature differences in normal years. Contrary to the assumption made by Le Quéré et al. to calculate $CO_2$ emissions in the residential sector, our data suggest that $CO_2$ emissions from gas use in buildings did not decrease in proportion to the intensity of COVID-19 confinements in each country (Figure 5). We do not have data for residential emissions from other fuels like oil and coal, the latter being important for heating in Poland. Nonetheless, our data suggest that temperature is indeed the dominant factor influencing gas-related emissions in the cold season. However, in six of the eight countries studied, we detected an additional (not temperature-related) reduction in $CO_2$ emissions that we tentatively attribute to COVID-19. Better interpretation of changes in 2020 will require more detailed spatial information about natural gas use patterns, associated with specific building types down to city or district levels, which is achievable through analyses of gas distribution network flows and smart-meters data.

**Methods**

Firstly, we collected data on daily natural gas flows towards distribution and final consumption in each country reported by operators to the ENTSO-G transparency platform ("Entso-G Transparency Platform," 2020). For the 8 countries considered, daily gas deliveries in kWh/day were downloaded from the ENTSO-G platform[16] for the period of 1-1-2016 to the latest available data. For most countries, the gas data was separated between distribution and final consumers, the latter one typically representing industry and power plants. For Belgium, final consumption was further divided in industrial use and power plants. For Germany, the data available on the ENTSO-G platform appeared to be incomplete. For this reason, daily gas consumption data was downloaded from online portals of GASPOOL (2020) and NetConnect Germany (NCG, 2020), the two main consortia of transmission system operators in Germany, and used instead of the ENTSO-G data. Since $CO_2$ emissions are highly correlated with the energy content of the gas, no further conversion was necessary. The so-called distribution sector in ENSTO-G corresponds to the consumption of gas in the built environment as a whole, and dominates the total gas consumption in each country. The data did not separate commercial, public and residential private buildings. For this sector, gas is mainly used for space heating and secondarily for cooking and hot water production. The distribution sector therefore shows a strong seasonal cycle with a peak consumption in winter, and fluctuations that relate to temperature anomalies. The power sector gas consumption shows very large variations linked to the total electricity demand, the weather-driven supply of solar and wind renewable electricity, the availability of nuclear, the relative distribution between coal, oil and gas that depends on market prices and climate policies. It is difficult to identify the impact of lockdown measures on gas consumption time series for the power sector, given the combined effects from other sources of electricity. The power sector is therefore excluded from our analysis. Additionally, we obtained daily consumption data for the City of Paris, (administrative jurisdiction, 2.18 million people [17]) based on local pipeline flows. Since there is almost no industry in the jurisdiction of Paris, all the use of natural gas can be assumed to come primarily from the residential sector.

Secondly, we collected daily index values summarizing government responses to the COVID-19, given by the Oxford Covid-19 Government Response Tracker (OxCGRT) [10]. The OxCGRT contains 19 indicators, grouped into "containment and closure policies", "economic policies", "health system policies" and "miscellaneous policies". Specific to the consumption in the distribution sector, we selected the "stay-at-home requirements". National containment and closure measures were adopted in Italy on March 9th with closure of all schools, most services and a large fraction of industrial activities. Similar measures followed in Spain on March 15th, in France on March 17th and in the UK on March 23rd. In other countries like the Netherland, Germany or Poland studied here, the confinement measures were not as severe. These measures have had a large impact on people's daily life and caused an economic downturn. A GDP reduction of more than 15% has been reported for Italy, Spain and France [18].

Thirdly, we collected the Google Community Mobility Reports representing how frequency of visits and duration of stays change against a baseline period (median value from the 5-week period Jan 3-Feb 6, 2020), at different places. For the gas consumption in the distribution sector, we selected the data for "residential".

Last, we collected national data on building types from a cross analysis of the databases of [11] and the Eurostat [19]. We extracted data on country population, final end uses of energy sources (power, gas, oil, coal, biomass) for commercial buildings, non-residential buildings, separated into energy used for space heating only and other end uses, average dwelling area, and total number of dwellings. These data are used to understand the country-specific characteristics of gas consumption changes with temperatures during the heating season.

Table 1: Result of the temperature dependent piecewise linear models applied to daily natural gas consumption data, as illustrated in Figure 2. For each cell of the table, the results for weekdays, Saturdays and Sundays are provided. For Spain, the consumption data could not be separated between distribution (buildings) and industry.

|  | Distribution | | | | Industry | | | |
| --- | --- | --- | --- | --- | --- | --- | --- | --- |
|  | $T_{crit}$ [°C] | Slope [mcm/day/°C] | Base [mcm/day] | Correlation | $T_{crit}$ [°C] | Slope [mcm/day/°C] | Base [mcm/day] | Correlation |
| Italy | 15.9 | 14.43 | 37.5 | 0.914 | 15.5 | 0.58 | 28.5 | 0.226 |
|  | 16.0 | 13.70 | 28.1 | 0.907 | 8.6 | -0.61 | 30.9 | 0.258 |
|  | 16.4 | 12.93 | 26.5 | 0.909 | 9.5 | -0.58 | 29.7 | 0.173 |
| Spain |  |  |  |  | 17.7 | 0.58 | 47.5 | 0.727 |
|  |  |  |  |  | 16.4 | -0.61 | 39.5 | 0.843 |
|  |  |  |  |  | 17.1 | -0.58 | 39.1 | 0.724 |
| France | 15.4 | 10.41 | 25.3 | 0.951 | 15.8 | 1.28 | 28.8 | 0.813 |
|  | 15.7 | 9.70 | 18.3 | 0.961 | 15.4 | 1.20 | 27.2 | 0.827 |
|  | 15.6 | 9.91 | 18.3 | 0.950 | 15.5 | 1.22 | 26.8 | 0.793 |
| Belgium | 14.6 | 2.50 | 8.0 | 0.938 | 15.1 | 0.70 | 12.6 | 0.824 |
|  | 14.4 | 2.40 | 6.4 | 0.945 | 15.0 | 0.63 | 11.6 | 0.802 |
|  | 14.7 | 2.36 | 6.4 | 0.935 | 15.3 | 0.66 | 11.6 | 0.778 |
| Netherlands | 13.5 | 6.66 | 16.1 | 0.924 | 15.2 | 0.85 | 22.1 | 0.289 |
|  | 13.4 | 6.48 | 13.3 | 0.925 | 14.6 | 0.49 | 23.4 | 0.242 |
|  | 13.6 | 6.30 | 12.9 | 0.920 | 16.0 | 0.54 | 23.3 | 0.335 |
| UK | 13.6 | 13.56 | 57.3 | 0.904 | 10.6 | 0.78 | 23.5 | 0.602 |
|  | 13.8 | 13.23 | 50.3 | 0.883 | 9.3 | 0.69 | 23.8 | 0.485 |
|  | 14.0 | 12.83 | 50.7 | 0.859 | 11.5 | 0.73 | 21.8 | 0.520 |
| Germany | 14.9 | 15.28 | 31.1 | 0.964 | 19.3 | 1.24 | 61.9 | 0.312 |
|  | 14.9 | 15.04 | 29.4 | 0.959 | 14.6 | 2.18 | 49.8 | 0.595 |
|  | 15.0 | 14.92 | 30.4 | 0.952 | 14.2 | 2.55 | 50.7 | 0.739 |
| Poland | 15.0 | 1.35 | 15.5 | 0.933 | 146.5 | -0.00 | 18.7 | 0.006 |
|  | 15.2 | 1.31 | 13.9 | 0.935 | -6.4 | -0.01 | 18.0 | -0.167 |
|  | 15.0 | 1.29 | 14.0 | 0.935 | 39.9 | 0.01 | 17.8 | 0.027 |

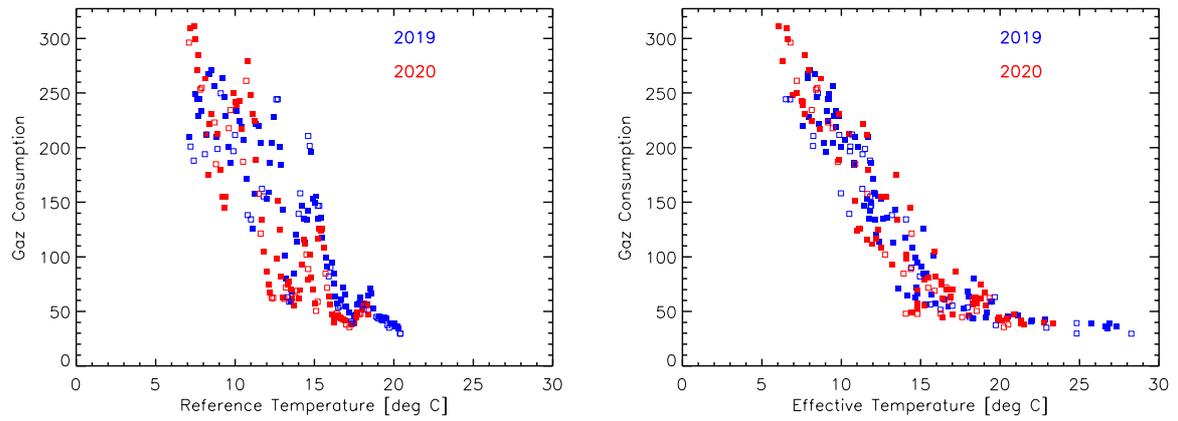

Figure 1. Daily gas consumption within the Paris city from March 1st to mid-June of 2019 and 2020. The gas consumption is shown as a function of the climatological temperature (a proxy of the date; left) and as a function of the effective temperature (right) i.e. normalized to remove the effect of daily temperature fluctuations (see text). Filled symbols are for the weekdays whereas the open symbols are for week-ends and holidays.

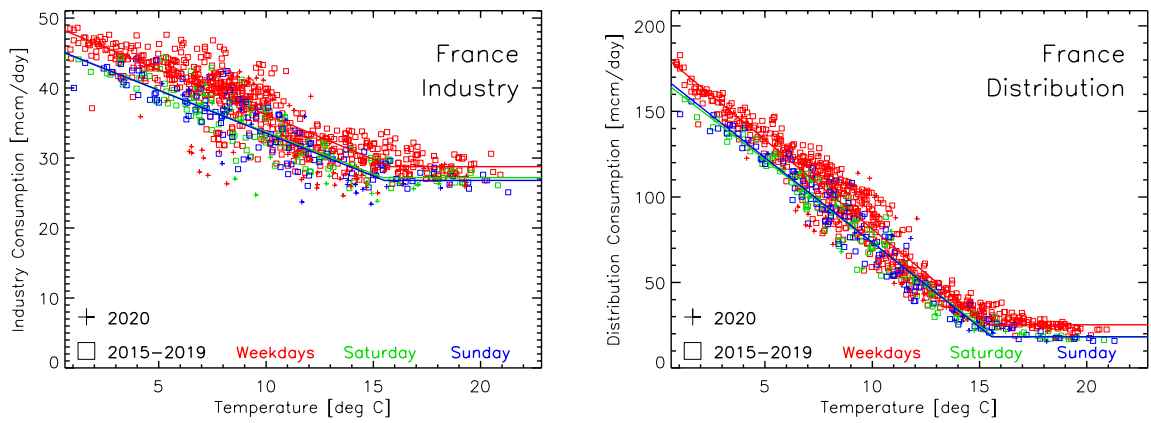

Figure 2. Daily gas consumption in France as a function of the effective temperature (see text) by the industry (left) and to the built environment, also called 'distribution' (right).

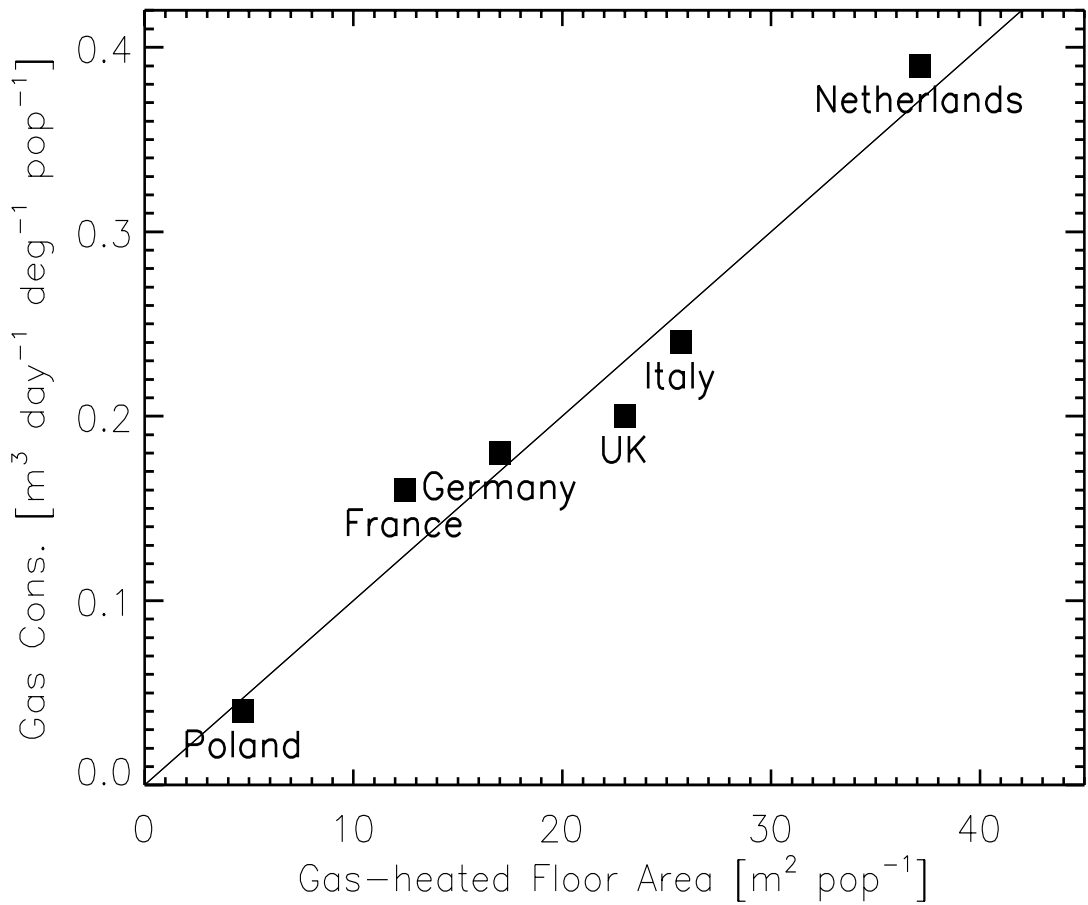

Figure 3. Regression between the temperature sensitivity of natural gas deliveries to the distribution network (the increase in per capita consumption per unit decrease of effective air temperature) and the per capita housing area, for houses heated by natural gas. The strong regression between the six countries suggests that building insulation and energy efficiency differences have small effects in controlling the sensitivity, as compared to the space available per capita.

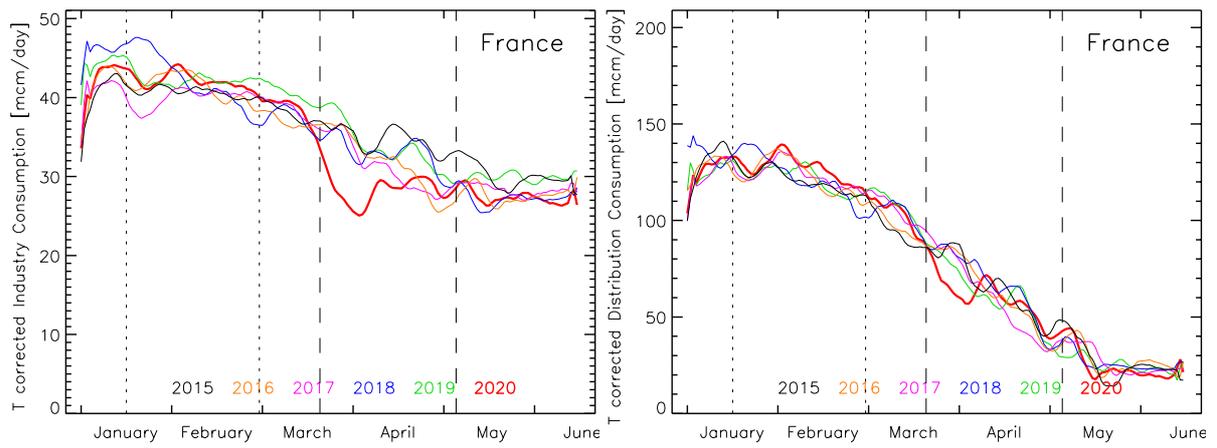

Figure 4. Time series of the daily natural gas consumption by the industry (left) and distribution to the built environment sector (right). Gas consumption values have been corrected for temperature anomalies using piecewise linear models, as in Figure 2, and time series are made smoother using the boxcar method over a period of 7 days.

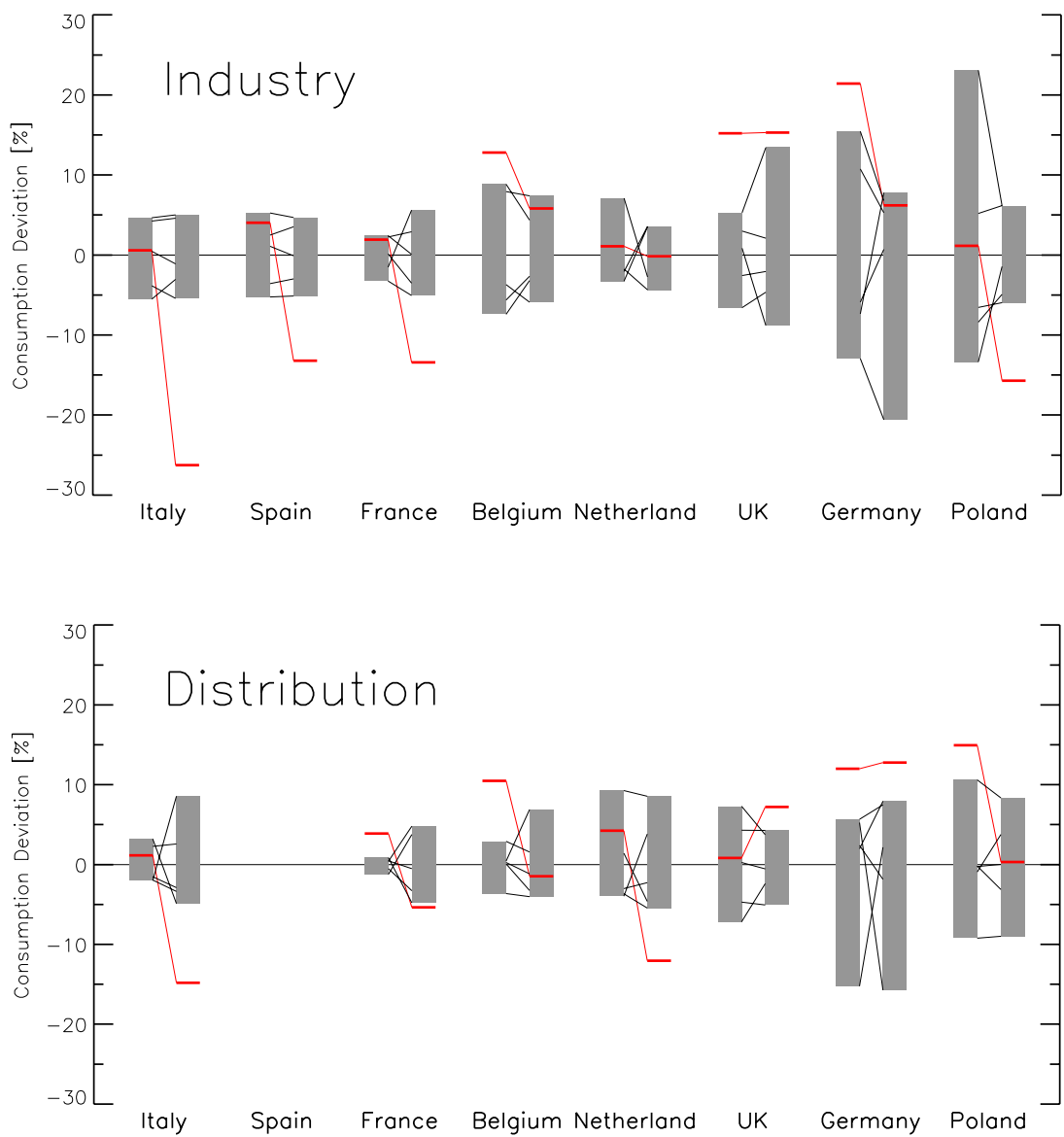

Figure 5. Deviations of the gas consumption in the built environment (distribution) in 2020 compared to the 2016-2019 mean. Two periods are considered: mid-January to end of February (left) and the first confinement period, which is country-specific (right). The bars indicate the range of the deviation from the period 2016-2019 and provide some indications of the inter-annual variations. For each country, the bar on the left is for the pre-confinement period, while the bar on the right is for the confinement period. The thin black lines indicate the variations for each individual year of the 2016-2019 period. The red lines show the 2020 values. All gas consumptions have been corrected for temperature anomalies.

**Supplementary material**

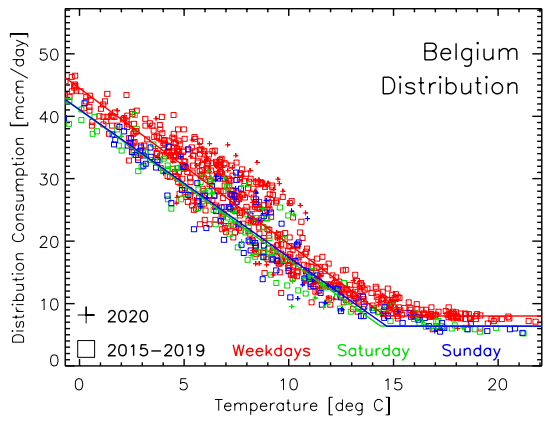
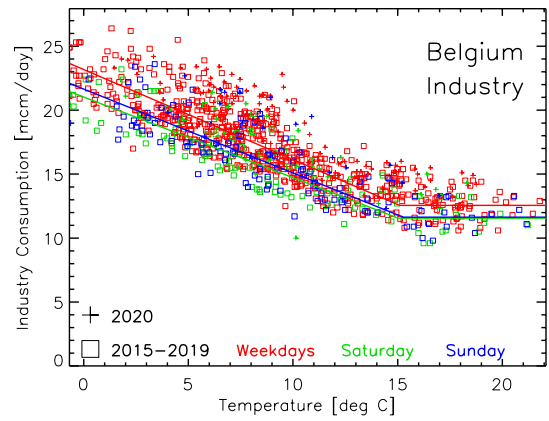
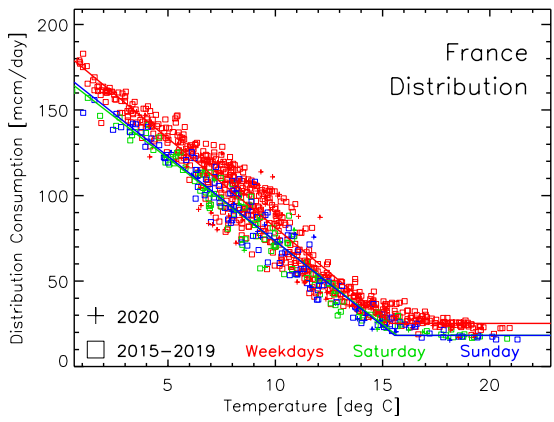
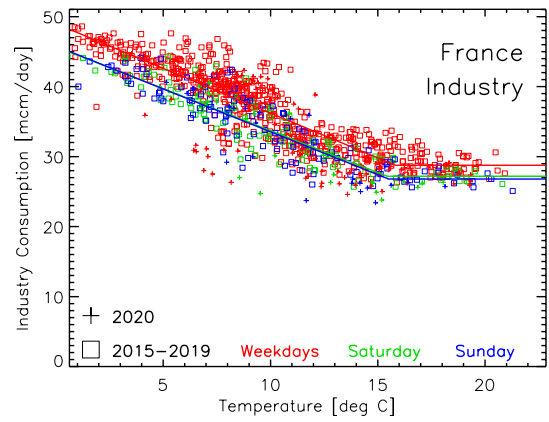
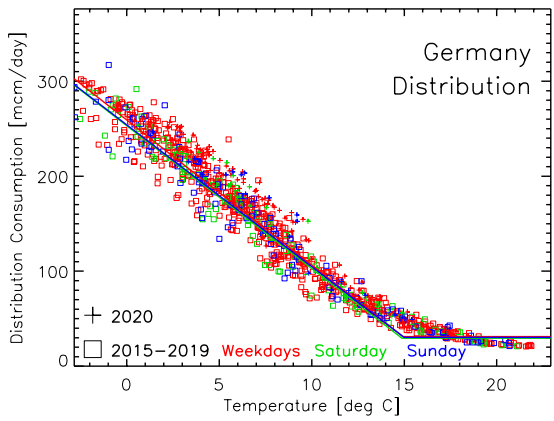
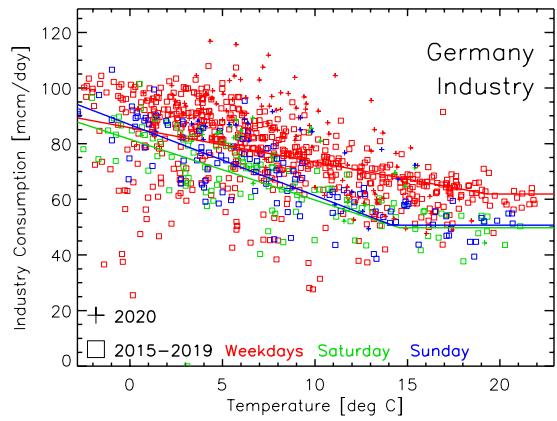

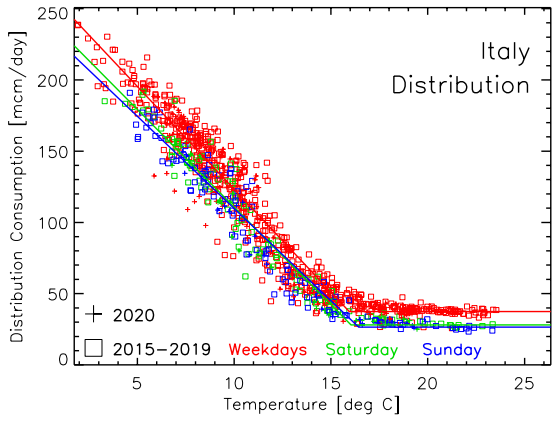
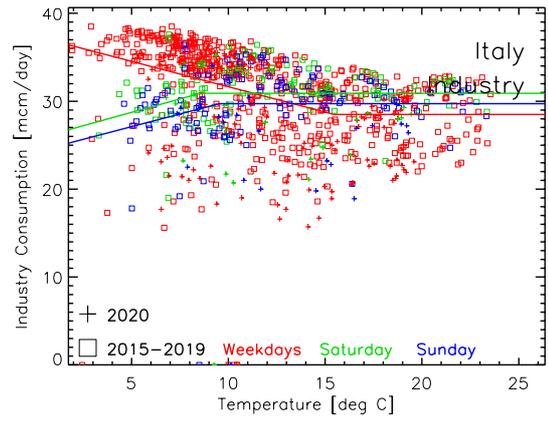
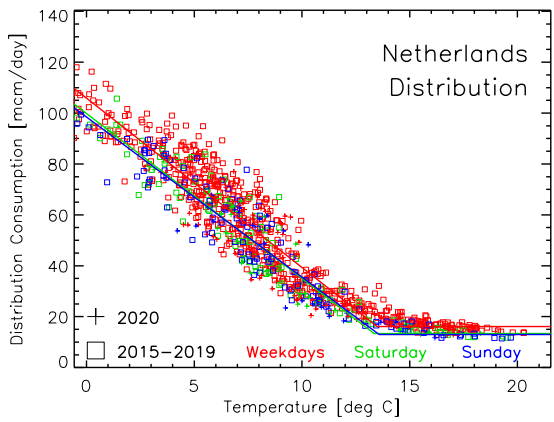
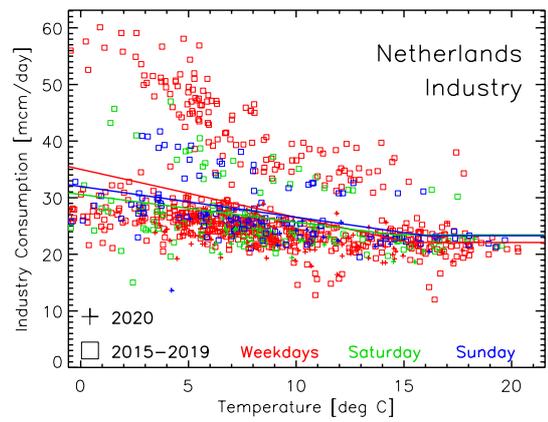
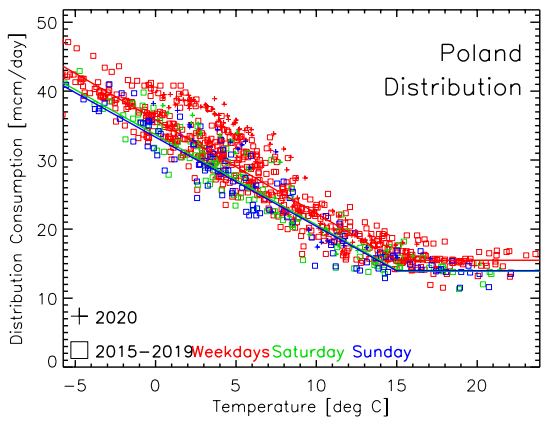
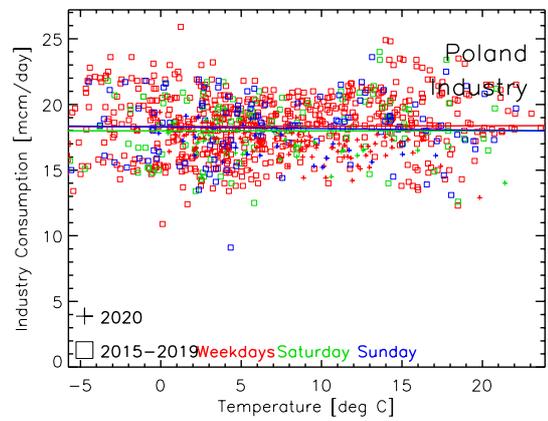

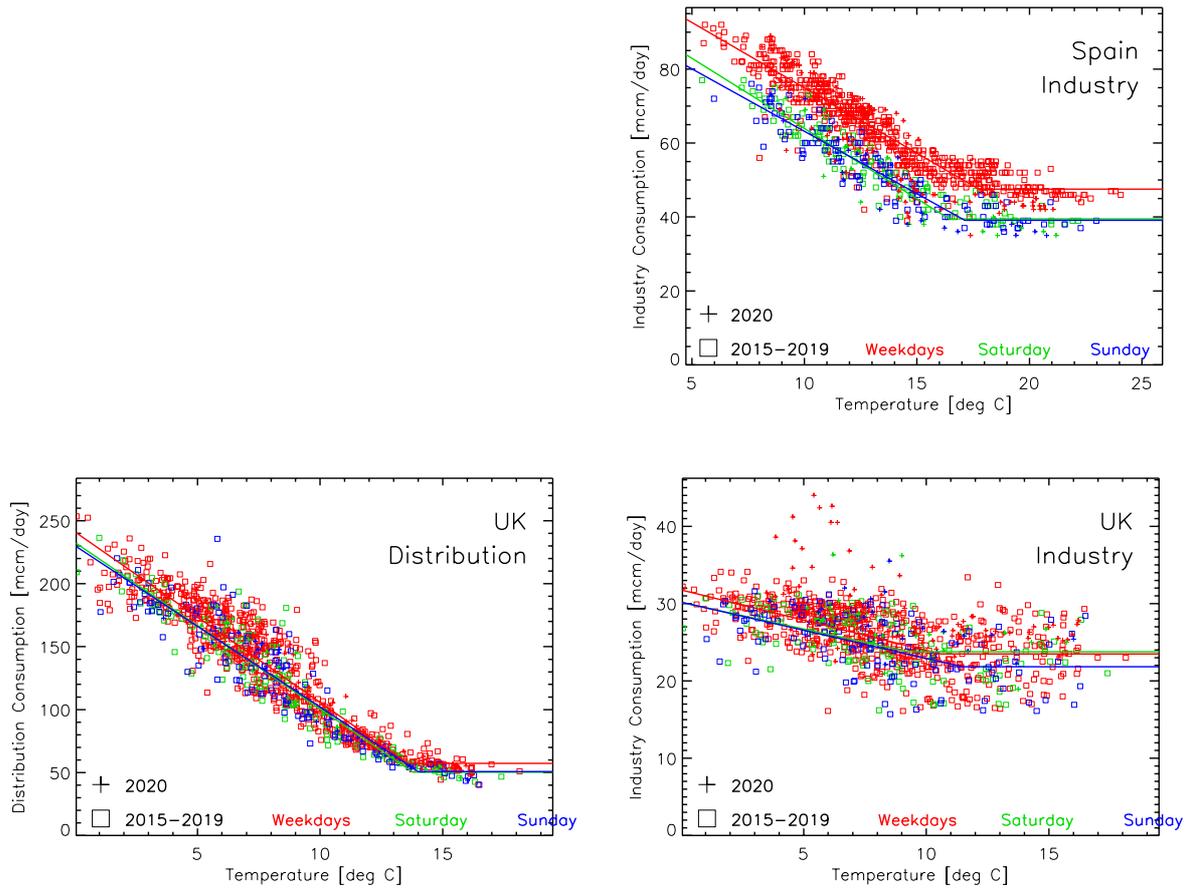

Figure S1. Regression of effective temperature (see text) and natural gas consumption in the built environment or distribution sector (left) and the industry sector (right) fitted by piecewise linear models. No data in Spain for the distribution sector that was not separated from the power sector.

France

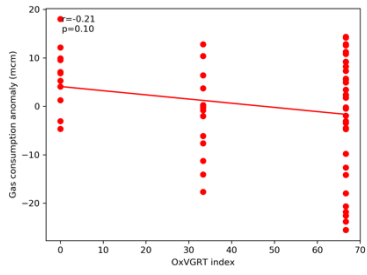 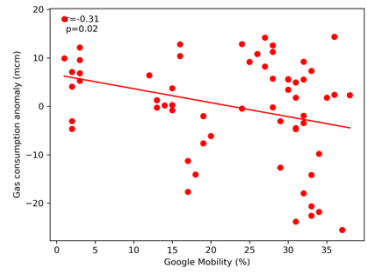

Belgium

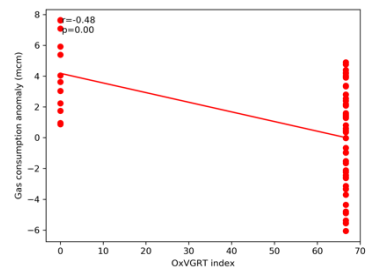 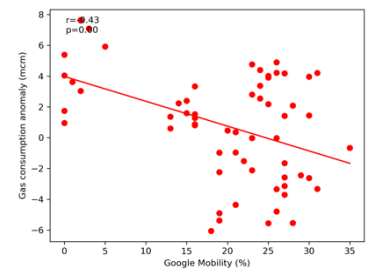

Italy

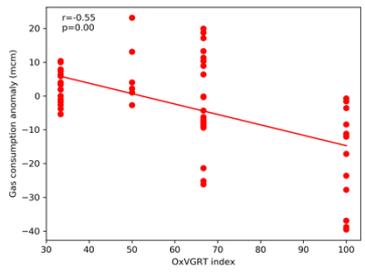 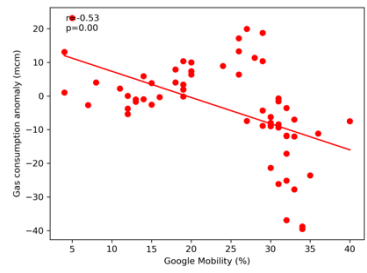

UK

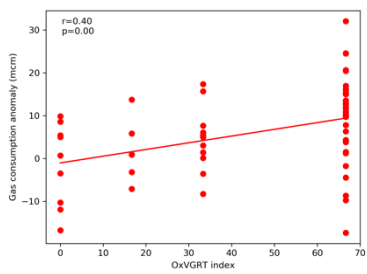 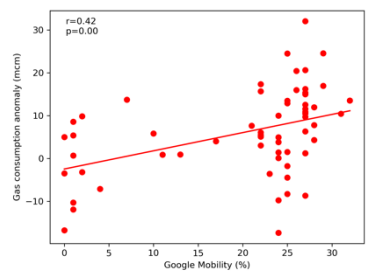

Germany

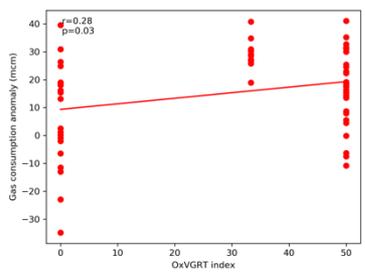 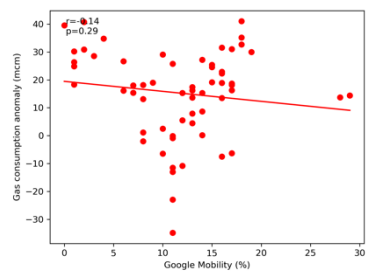

Poland

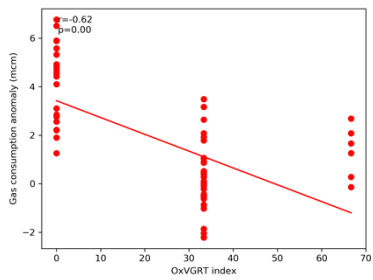
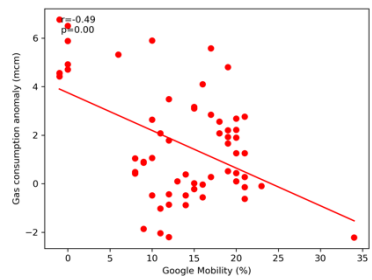

The Netherlands

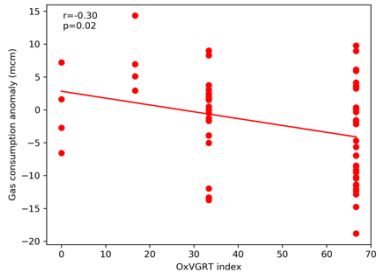
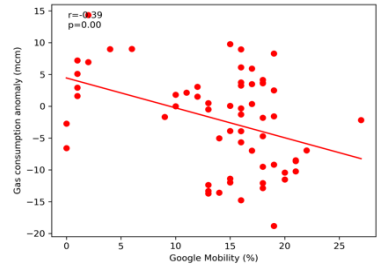

Figure S2. Regression of temperature corrected daily natural gas consumption during lockdowns and OxCGRT confinement severity indexes (discrete values with 0% being no government response) on the right, and the Google mobility index of home presence anomaly each day with respect to a reference period of Jan 3 to Feb 6, 2020) on the left.